# Ultrafast near-field imaging of exciton-polariton dynamics in WSe$_2$ waveguides at room temperature

Michael Mrejen, Lena Yadgarov, Assaf Levanon, Haim Suchowski[§]

School of Physics and Astronomy, Faculty of Exact Sciences
Tel Aviv University, Tel Aviv 69978, Israel

[§]Correspondent author: haimsu@post.tau.ac.il

Van der Waals (vdW) materials, weakly bound layered compounds, have received enormous interest as they offer a malleable playground for a wide range of physical properties in thermal, electronic and optical devices. In particular, owing to their inherent deep subwavelength light confinement, they support a variety of light-matter interactions phenomena such as plasmons, light emission, excitons and phonons conveyed as polaritonic modes (*1, 2*). Specifically, semiconductor vdW materials (*3*) such as tungsten diselenide (WSe$_2$) are particularly attractive for photonic and quantum integrated technologies since they sustain Visible- Near-Infrared (VIS-NIR) exciton-polariton (EP) modes at room temperature (*1, 2, 4-6*). In the quest to unravel the underlying physics of these intriguing phenomena, advanced subdiffraction imaging techniques such as scanning near field optical microscope (SNOM) (*7, 8*) has provided valuable insights on the nature of the EP coupling mechanism to the waveguide modes sustained in vdW materials *(5, 6)*. While most of these works focused on the steady state of the EP, the spatio-temporal dynamics of EP formation, happening in the sub-picosecond regime, remains largely unexplored. Hence, a direct

**imaging at the femtosecond-nanometric scale of the EP evolution is critical to the understanding of these coupled light-matter states. Here we report for the first time the ultrafast and deep-subwavelength imaging of exciton-polariton formation and propagation in WSe$_2$ waveguides. Our method, based on a novel ultrafast intra-pulse pump-probe near-field imaging, allows to directly visualize and quantify the EP time evolution at room temperature. More specifically, and in agreement with our time-dependent model, we directly observe a significantly slow EP wave packet group velocity of $v_g \sim 0.017c$ which is attributed to the bandgap renormalization originating in the light coupling near the exciton transition. These findings suggest that vdW based materials could be used not only as a platform for valleytronics as it has been shown so far but also for slow light with applications in light storage for memory, enhanced optical nonlinearity, sensing and more. Moreover, our experimental method paves the way for in-situ ultrafast coherent control of other polaritons in low dimensions materials as well as for extreme spatio-temporal imaging of condensed matter systems.**

In recent years, vdW materials, such as graphene and hexagonal boron nitride (hBN), have emerged as a new material system supporting various types of mixed modes of light and crystal polarization, a.k.a polaritons, embodied as plasmon polariton and phonon polariton with unique properties (*9-13*). On the other hand transition metal dichalcogenides (TMDs) with chemical formula MX$_2$ (M = Mo, W; X = S, Se, Te) are vdW semiconductors with sizeable bandgaps and strongly bound excitons (*3, 14*). These excitons can couple with photons, provided the mean to bridge the momentum gap, and form exciton polaritons (*1, 2*) which have been observed in the visible-Near IR spectral region at ambient conditions.

Indeed, far-field optical studies of TMDs embedded in microcavities have captured the spectroscopic signatures of strongly coupled cavity EPs (*15-18*). More recently, the light-exciton coupling has been demonstrated to be achievable via waveguide modes in subwavelength vdW materials structures *(4-6)*. The real-space characteristics (for example, propagation, confinement and interference) of the TMDs EP have been addressed only very recently with near-field imaging where their wave-like behavior has been unraveled and characterized (*5, 6*). While these groundbreaking studies provide important insights on the steady state behavior of the EP, the temporal dynamics of their formation has remained unaddressed to date.

In parallel, in the past decade much progress has been achieved in deep subwavelength imaging using labeled techniques such as PALM, STED and STORM (*19-21*), yet label-free optical microscopy remains plagued by the diffraction limit and therefore is unable to resolve details on the nanometer scale. While the Near-field Optical Scanning Microscope (SNOM) breaks this barrier, achieving up to 3 orders of magnitude higher resolution than conventional microscopy in the optical regime, it fails to provide extreme temporal resolution. Previous reports of pump-probe spectroscopy in the near-field were either based on aperture SNOM with spatial resolution limited to > 100 nm or have used relatively narrow ultrafast sources with probe pulses width ranging from 220 fs to 5 ps (*22, 23*). Although these approaches have been able to probe up to sub-picosecond dynamics, they are unsuitable to resolve extreme ultrafast phenomena at the sub 100 fs regime and in particular the evolution of photo-induced carriers dynamics in materials. Here we report the direct imaging of the propagation of a guided exciton-polariton wave in $WSe_2$ slabs in the VIS-NIR spectral region and at room temperature with unprecedented spatio-temporal resolution of 50 nm and 66 fs. Our novel ultrafast broadband near-field apparatus overcomes the aforementioned limitations and enables the observation of a significant slowing down of the EP group velocity down to $v_g \sim 0.017c$ in a uniform vdW material. This observation constitutes, to the best of our knowledge, the first direct observation at the nanoscale in a TMD of the modification of the dielectric function (a.k.a. 'renormalization') introduced by an intense laser excitation around the exciton resonance.

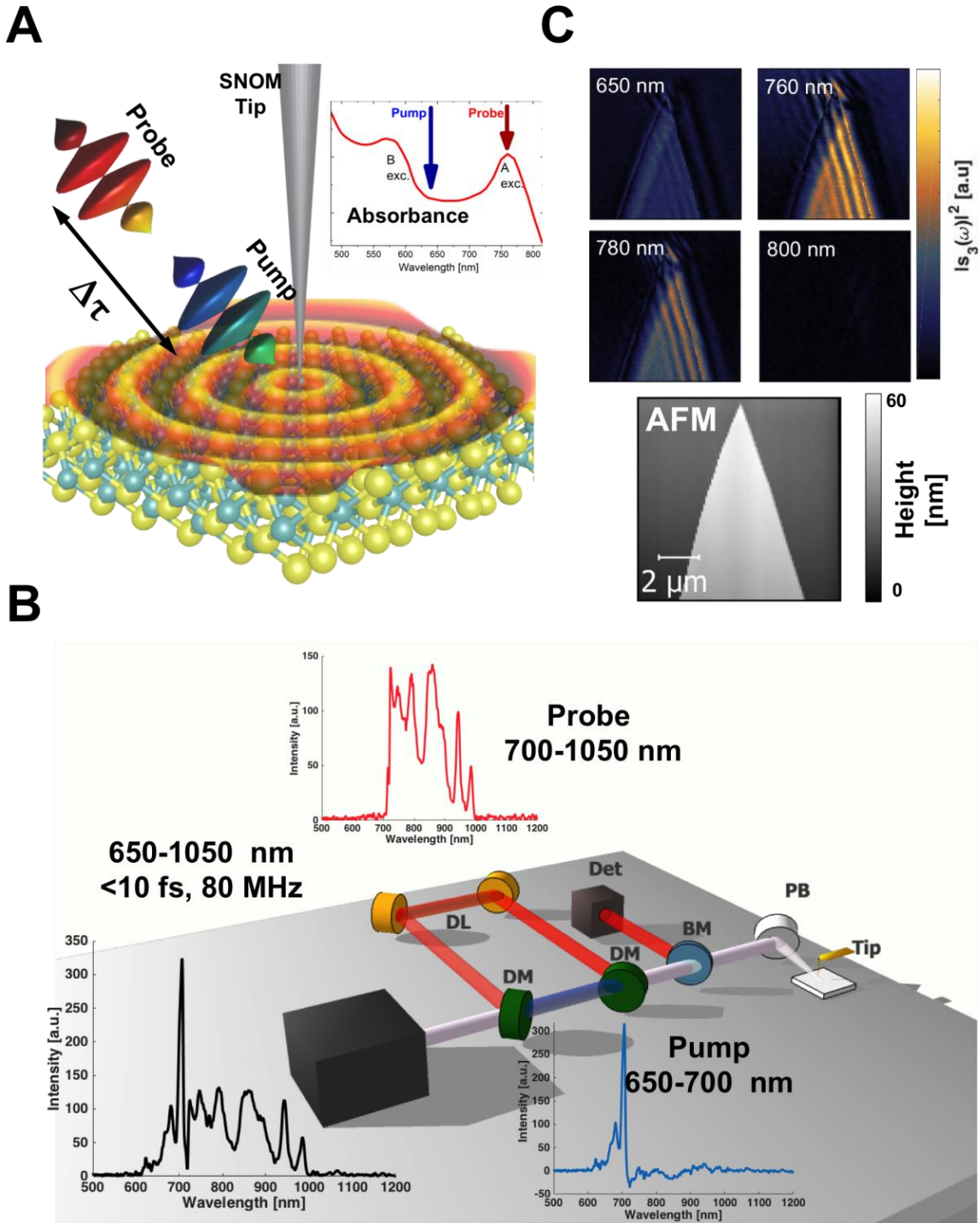

Figure 1- Near-field Ultrafast and broadband Pump-probe of EP in WSe$_2$. (A) Schematic of the ultrafast pump-probe EP excitation and detection. The pump pulse impinges on the tip apex which localizes the pulse energy and excite an excition-polarition wave packet which propagates on the WSe$_2$ slab. The wave packet is reflected at the boundaries of the flake and within its thickness and ultimately scattered back by the tip. (B) Schematic of the experimental near field pump-probe apparatus. The ultrafast sub 10 fs pulses Ti;Sapphire source, spanning a broad spectrum of 650-1050 nm, is split by a dichroic mirror (DM) into two pulses: pump encompassing the higher energies of 650-700 nm and the probe from 700-1050 nm. The probe is delayed with respect to the pump by steps of 66 fs. The two pulses are recombined after the delay

In order to probe the spatio-temporal behavior of the exciton-polariton formation, we have fabricated WSe$_2$ micron scale slabs with thicknesses ranging from atomic monolayer to several tens of monoloayers (see Methods for fabrication). These slabs are characterized by two pronounced excitonic resonances A and B at 1.67 eV and 2.18 eV respectively (Inset Fig 1A). We then employed a unique probe-probe near-field microscope where an ultrafast laser (sub-10 fs) spanning a spectrum from 650 nm to 1050 nm has been combined with an apertureless SNOM as seen in Fig 1A. The laser beam was split into two beams, the pump, on one hand, from 650 nm to 700 nm and the probe on another hand, from 700 nm to 1050 nm, encompassing the A exciton transition. The probe was significantly attenuated and delayed with respect to the pump using a delay line and the recombined beams were focused on the SNOM tip (for more details in regarding the pump-probe synchronization see the Supplementary Material). WSe$_2$ slabs are scanned under the tip, which by strongly localizing the incoming electrical field, bridges the momentum gap necessary to the launching of EP (*1, 2*) (Fig 1B). The tip scattered light is collected, the pump beam is filtered out at the detector and the probe signal is spectrally analyzed. We first conducted probe only scans to spectrally resolve the steady state response of the flake and observed an enhanced near-field signal at 760 nm (with 10 nm bandwidth), in the vicinity of the A transition (Fig 1C). The SNOM image clearly exhibits the interference fringes characteristic of the exciton-polariton wave interacting with the incoming and outgoing light and are in agreement with recent reports (*5, 6*) (See Supplementary information). We then carried out pump-probe experiments and chose to probe the system at this wavelength for optimal response. In the pump-probe experiments the probe is delayed with respect to the pump by using a precisely controlled delay line (by steps of 10 µm, $\Delta\tau$~66 fs, total time span of 1.1 ps) and for each delay the flakes are scanned under the SNOM tip covering areas of 8 µm x 8 µm with a pixel size of 50 nm (See SOM for additional information). In Fig 2A we present a selection of the time near-field snapshots collected (all the frames collected with temporal resolution of 66 femtosecond are supplied as a supplementary movie).

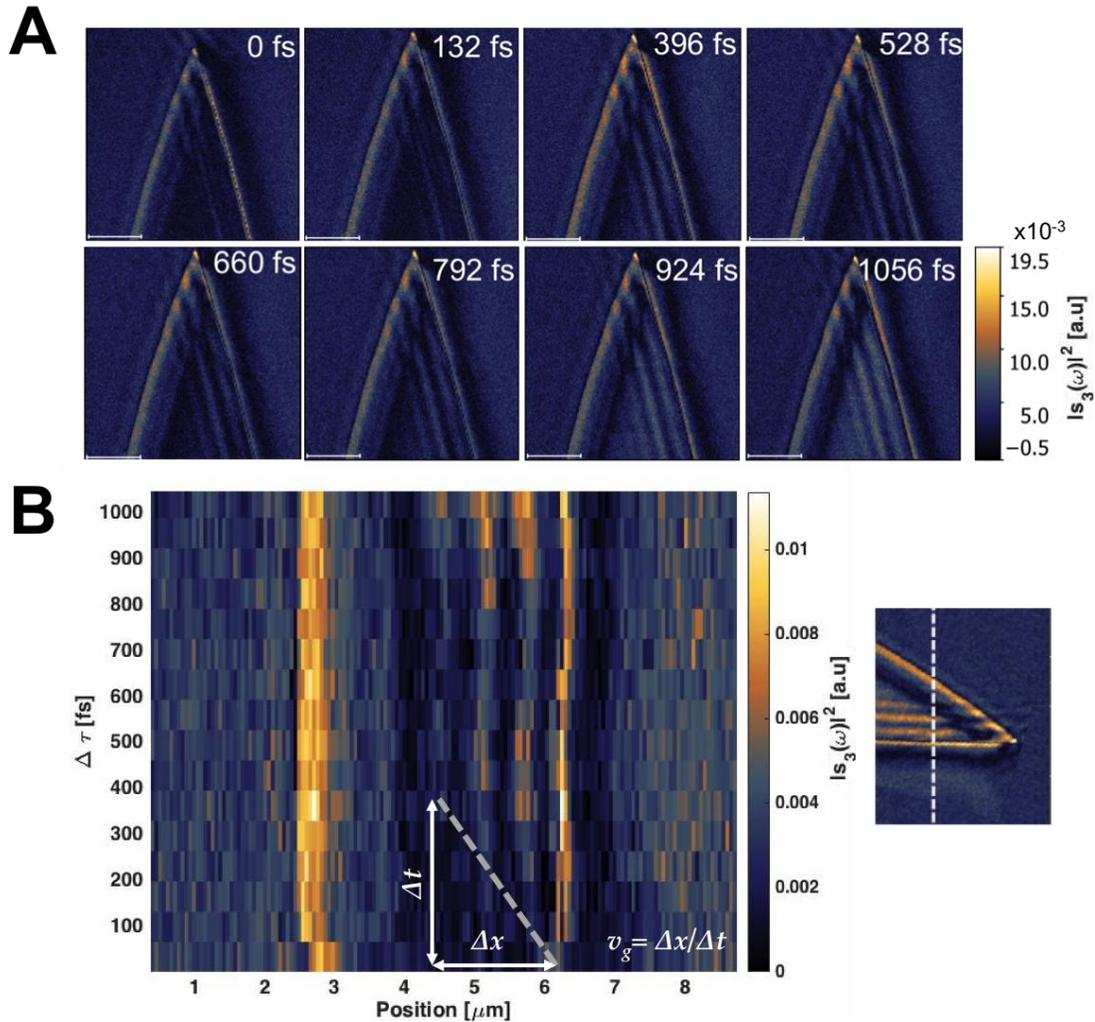

Figure 2 – Pump-probe near-field images showing the evolution of EP wave packet in WSe$_2$ (A) Selected snapshots of the SNOM images collected at different time delays for a probe wavelength of 760±5 nm. The SNOM signal depicted is $|S_3(\omega)|^2$, the intensity of the scattered signal demodulated at the third harmonic of the tip resonance frequency $\omega$. A clear emergence of the interference fringes is observed across the flake as a function of the time delay. The thickness of the WSe$_2$ flake is measured to be 60 nm, ~100 layers, discussion and SNOM data for flakes ranging from monolayer to over 90 nm can be found in the main text and in the Supplementary material. (B) Two dimensional map of the SNOM signal as a function of the position across the flake along the white dotted line (right inset) and as a function of the pump probe time delay. We observe the consecutive appearance of interference fringes as the position goes inward inside the flake and as the time delay increases. As we show further in this letter, this is the expression of the EP wave packet excited by the tip, that propagates and bounces back from the nearby flakes' boundaries at a group velocity of $v_g$. Therefore $v_g$ can be retrieved by analyzing the time elapsed till the appearance of a particular fringe with respect to its position relative to the boundary. The grey dotted line is a guide for the eye to show the experimental retrieval of $v_g$ which we found to be equal to $v_g = 4.7 \pm 0.5 \times 10^6$ m/s.

These time snapshots distinctly reveal a dynamical change across the slab with the emergence of the typical interference fringes over the time span of the measurements. Qualitative visualization of this change is obtained by plotting the near-field scattered signal along a cross-section of the flake (depicted as the dashed white line in the inset of Fig 2B) versus the time delay between the pump and the probe. The temporal appearance of the interference fringes is striking and we observe their consecutive appearance with an unprecedented spatio-temporal resolution of 50 nm and 66 fs, first close to the flake's boundaries and then as they move inward.

In order to gain an understanding of the underlying physical mechanisms and as an extension of recent reports (*1, 5, 6*), we modeled the exciton-polariton time-dependent excitation, propagation and detection taking into account the different contributions at different time delays to the collected signal depicted in Fig. 3A. The incoming wave is first scattered back to the detector by the tip (S1 in Fig. 3A). At the same time it excites, via the tip, the exciton-polariton mode that is guided by the flake slab-like waveguide and is back reflected toward the tip by the nearby boundaries. Those reflections, once reaching the tip are scattered toward the detector. Finally, the light trapped in the thickness of the slab also contributes to the overall scattered signal as it exits the slab. All these contributions amplitudes add up with their respective phases at the detector and give rise to the interference patterns that we observe (Fig 3B).

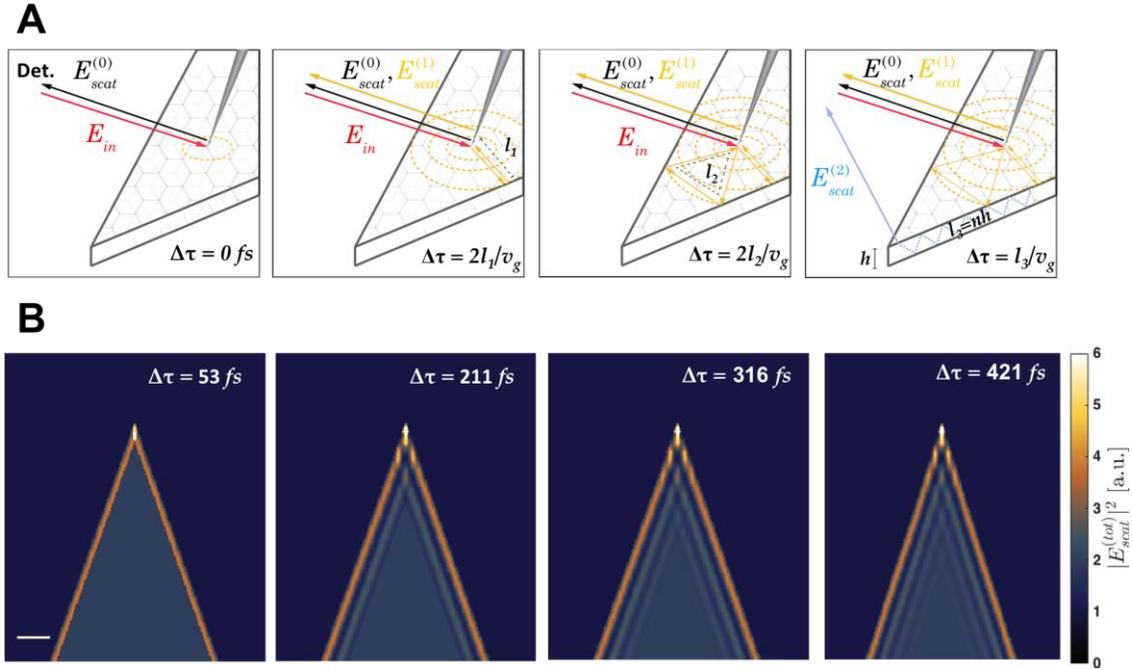

**Figure 3 - Modeling of the near-field exciton-polariton spatio-temporal propagation (A)** Schematic of the different optical paths that contributes to the SNOM signal at different time delays according to the velocity of the wave packet excited at the tip location. At t=0, the EP wave packet (depicted as yellow dotted circle) is launched at the tip by the incoming field $E_{in}$ and start to propagate along the flake. At this point only direct scattering by the tip $E^{(0)}_{scat}$ contributes to the near-field signal collected at the detector (Det.). Further on, as the EP wave packet propagates along the flake, it bounces back from a nearby boundary towards the tip and, at the adequate time delay that corresponds to the round trip from the tip to the boundary ($2l_1$), reaches it and contributes an additional component, $E^{(1)}_{scat}$ to the scattered near-field amplitude. Further secondary (third panel) and tertiary (not depicted) back reflections are also taken into account in accordance to previous models (*1*). An additional contribution, $E^{(2)}_{scat}$, accounts for the light trapped in the thickness of the flake and that bounces back and forth *n* times till it exits the flake toward the detector (depicted as light cyan dotted line in the last panel). All these contributions add up with different phases and at different times, if at all, according to the time delay and the relative position of the tip and the flakes' boundaries. **(B)** Numerical results of the described model depicted as the total near-field scattered intensity. We observe that the model predicts the consecutive appearance of one to four fringes in about 420 fs corresponding to a group velocity of $v_g = 5.2 \times 10^6$ m/s, in a good quantitative agreement with the experimentally retrieved $v_g$. Furthermore this quantitative agreement yields a propagation loss of $\gamma = 2.64$ μm$^{-1}$. This relatively high loss figure suggests that the pump pulse heavily dopes the WSe$_2$. white scale bar: 1μm.

The results of our time-dependent model presented in Fig. 3B reveals that the pattern we observe are constructive interferences of the EP wave packets that propagate inside the WSe$_2$ slab waveguide. Therefore this direct spatio-temporal imaging of the EP wave packet propagation allows us to retrieve its group velocity, yielding a remarkably low velocity of

$v_g = 5.2 \times 10^6 \, m/s$, two orders of magnitude slower than the speed of light in vacuum. This surprisingly slow velocity is in agreement with previous similar obervations in the far field (*24*) and is a manifestation of the significant variation of the group velocity around the discrete A exciton resonance due to the renormalization caused by exciton-light field coupling (*25*). These measurements therefore constitute, to the best of our knowledge, a first of its kind.

We further conducted measurements on various slab geometries and illumination configurations (angle of incidence and polarization). We found that these fringes originates from wave packets they are back reflected toward the tip by the boundaries, their appearance thus strongly relies on the geometry of the flakes and the relative position of the tip to the flake's boundaries. Moreover the asymmetry observed in the measured fringes originates from the asymmetric excitation, collection and detection dictated by our apparatus geometry (see Supplementary material). Therefore, such geometrical considerations have crucial importance in designing future devices based on vdW and hetro-structures materials. We also would like to mention that several experiments have been performed on single-layer and few layered flakes, and did not show such interference fringes or evolution, which support the claim that such evolution is related to the propagation mode of such materials.

We emphasize that compared to slow light schemes in atomic systems (*26-29*) or in periodic dielectric structures (*29-35*) which require bulk-material resonances, disorder or periodic structuring, nanoscale guiding structures in 2D materials inherently feature slow waves mediated by surface plasmon polaritons in graphene (SPPs) or surface phonon polaritons in hBN. Beyond their ability to confine light beyond the diffraction limit, such platforms present the advantage of allowing the control of the speed of the guided waves via electrical doping. However, the aforementioned 2D material platforms span the mid-IR spectral region therefore excluding their use for slow light in the VIS-NIR region. Our observation therefore is pointing toward a slow light 2D platform in this important spectral region from a technological and application point of view.

In conclusion, in this Letter we have reported, for the first time, the direct imaging of the formation and propagation of exciton-polariton in WSe$_2$ waveguides at room temperature. Our unique ultrafast intra-pulse pump-probe near-field imaging allows to directly visualize and quantify the EP time evolution with unmatched spatio-temporal resolution of sub 70 fs and 50 nm at NIR-VIS Wavelength regime. We report the observation of a strikingly slow EP wave packet group velocity of $v_g \sim 0.017c$, two order of magnitudes slower than the speed of light. This remarkable slowing down of the group velocity is originating in dispersion renormalization around the A-exciton resonance. We have also found good agreement between our experimental observation and our time-dependent model that includes the contributions from both the waveguide mode and the slab geometry. These findings suggest that vdW based materials could be used not only as a platform for valleytronics as it has been extensively shown in the recent years but also for slow light with applications in light storage for memory, enhanced optical nonlinearity and sensing and more. Finally our experimental approach open the way for extreme spatio-temporal imaging and control of light-matter interaction.

**Acknowledgement**

This research has been supported by the European Research Council (ERC), under the project MIRAGE 20-15.